\documentclass[%
 reprint,
 amsmath,amssymb,
 aps,
]{revtex4-2}
\usepackage[colorlinks=true,urlcolor=blue,linkcolor=blue,citecolor=blue]{hyperref}
\usepackage{amsmath} 
\usepackage{amssymb}
\usepackage{epsf}
\usepackage{color}
\usepackage{dcolumn}
\usepackage{bm}

\usepackage{graphicx}
\graphicspath{ {./images/} }

\begin{document}

\title{A new quantum hydrodynamic description of ferroelectricity in spiral magnets}%

\author{Mariya Iv. Trukhanova$^{1,2}$}
\email{trukhanova@physics.msu.ru}
\author{Pavel Andreev$^{1}$}
\email{andreevpa@physics.msu.ru}
\author{Yuri N. Obukhov$^{2}$}
\email{obukhov@ibrae.ac.ru}

\affiliation{$^{1}$~Faculty of Physics,
Lomonosov Moscow State University, Leninskie Gory 1, 119991 Moscow, Russia  \vspace{1em} \\
$^{2}$~Theoretical Physics Laboratory, Nuclear Safety Institute, Russian Academy of Sciences,
B. Tulskaya 52, 115191 Moscow, Russia}

\begin{abstract}
The strong coupling between magnetism and ferroelectricity was found in rare earth manganites, where the electric polarization could be induced by special magnetic ordering. There is no theoretical model that would allow us to study the static and dynamic properties of electric polarization in strongly correlated magnetic dielectrics. In the presented research, we have taken the main step towards the construction of such a fundamental model, and made a direct connection between the microscopic Katsura-Nagaosa-Balatsky theory and Mostovoy's phenomenological model for magnetically induced polarization. A novel description of the ferroelectricity of spin origin is proposed within the framework of the many-particle quantum hydrodynamics method. It is applied to the study of cells of magnetic ions, where the electric dipole moment is proportional to the vector product of spins. Our approach is based on the many-particle Pauli equation, where the influence of an external magnetic field is considered. We define the electric dipole moment operator of the ion cell and introduce the macroscopic polarization as the quantum mechanical average of that operator. We formulate a model for the description of nonequilibrium polarization and derive a new polarization evolution equation. The polarization switching in ferroelectric magnets with the spiral spin-density-wave state is considered, and we demonstrate that the proposed model yields known results and can predict novel effects. The dynamic magnetoelectric effect can be investigated by employing this novel equation to study the evolution of polarization.
\end{abstract}
\maketitle
\section{Introduction}
A characteristic property of magnetoelectric materials is the occurrence of an electric polarization in an external magnetic field, or magnetization in an external electric field. In 1956, Landau and Lifshitz introduced a more precise concept of magnetoelectric materials as media whose symmetry allows the existence of a linear magnetoelectric effect \cite{LL}. On the other hand, the magnetoelectric effect was predicted by Dzyaloshinskii in 1959 for Cr$_2$O$_3$ in weak magnetic fields for an unstrained antiferromagnetic structure \cite{Dzyaloshinski1960OnTM}. In the next year, Astrov experimentally confirmed this effect when he measured the magnetization induced by the electric field \cite{Astrov}. In 1961, Folen and colleagues \cite{Rado, Wiegelmann} obtained the polarization caused by a magnetic field, see also Refs. \cite{Fiebig,Pyatakov,Pyatakov2}. There are materials called multiferroics, which, even in the absence of external electric and magnetic fields, can unite the spontaneous ordering of magnetic and electric moments \cite{Khomskii, Lines}. The study of the magnetoelectric coupling properties in multiferroics is important for controlling the magnetic memory and studying biological systems. For example, the magnetoelectric effect can be used to wirelessly stimulate certain areas deep in the brain \cite{Kopyl, Koz, Chen:2023} or magnetoelectric materials can be used for tissue engineering \cite{Surmenev}.

There are different microscopic mechanisms of ferroelectricity and corresponding types of multiferroics. In type-I multiferroics, ferroelectricity and magnetism have different sources. In these multiferroics, ferroelectricity appears due to lone pairs, charge ordering, or can have a geometric origin \cite{Khomskii}. For the past few years, much attention has been devoted to the study of type-II multiferroics. In these materials, the nontrivial spin structure of the crystal lattice causes electric dipole polarization \cite{Aken, Jia, Arima}. In rare-earth manganites with an orthorhombically distorted perovskite structure, ferroelectricity often appears due to the antisymmetric exchange interaction \cite{Hig, Kim} and the helical spin order. But, in general, different types of spin order can potentially break the inversion symmetry of space, leading to the induction of polarization.

Type-II multiferroics with strong magnetoelectric coupling were subsequently discovered in the laboratory. Kimura experimentally showed \cite{Kimura} that the rare-earth manganite TbMnO$_3$ is a typical example of a material that has an orthorhombically distorted structure and can exhibit a large magnetoelectric response \cite{Kim, Kimura2}. The magnetic moments of positively charged manganese ions Mn$^{3+}$ in the sample pass to an antiferromagnetic state at $T_n \sim 41$ K. At the temperature $T_c\sim 28$ K, a cycloidal spiral magnetic structure of manganese ions appears that leads to the ferroelectric order with spontaneous electric polarization \cite{Kajimoto, Kenzelmann}. Despite the fact that the electric polarization magnitude in the TbMnO$_3$ material takes rather small values, it can be easily controlled by external homogeneous magnetic fields \cite{Kimura3}. The external magnetic field of 4-8 T strength can switch the polarization, which was induced along the $c$-axis of the crystal, by 90$^o$ degrees if the Mn$^{3+}$ ions have an elliptically modulated noncollinear helical structure and are oriented in the $bc$ plane \cite{Tokura,Zvezdin}.

There are three main mechanisms for the occurrence of spin origin ferroelectricity in type-II multiferroics: symmetric exchange interactions or exchange-striction mechanism \cite{Kobayashi}, antisymmetric exchange interaction, leading to the appearance of polarization under the action of spin-orbit coupling and described by the inverse Dzialoshinskii-Moriya model \cite{Katsura,Mostovoy,Sergienko}, and spin-dependent $p$-$d$ hybridization \cite{Arima,Seki}. The theoretical description of the magnetoelectric effect is a nontrivial task since, at the microscopic level, there are many factors influencing the origin of the electric dipole moments of crystal lattice elements and the polarization of the sample. The phenomenological approach of Ginzburg and Landau is often used to describe magnetoelectricity. In Ref. \cite{Harris}, the Landau decomposition for magnetoelectric coupling in incommensurate ferroelectric magnets was demonstrated in terms of complex-valued magnetic order parameters. For inhomogeneous helical ferroelectric magnets, M. Mostovoy obtained an effective phenomenological model based on the Ginzburg-Landau approach that describes their thermodynamics and behavior in a constant magnetic field \cite{Mostovoy,Mostovoy2}. The microscopic origin of the electric polarization, induced by noncollinear magnetic order, was investigated in Ref. \cite{PhysRevLett.127.187601}. The electric dipole moment of the bond was obtained using all possible combinations of position operators and transfer integrals in the basis of Kramers doublet states.   

The antisymmetric exchange interaction energy may be derived from the spin-orbit correction to the Anderson superexchange \cite{Anderson}. The polarization of a bond, which is proportional to the vector product of spins, was obtained using the spin-current model and the Dzialoszynskii-Moriya inverse interaction model \cite{Moria}. The spin-current model was proposed by Katsura, Nagaoshi and Balatsky \cite{Katsura,Hu}. In their approach, the indirect exchange interaction between two neighboring positively charged magnetic ions is carried out through a negatively charged ion-ligand (e.g., oxygen, O$^{2-}$). The electric dipole moment is induced in the cluster of these ions as a result of the covalent bonding of $d$- and $p$- electrons. It is considered that a spin current flows between two non-collinear electron spins on the $d$-orbitals of magnetic ions. The Dzyaloshinskii-Moriya inverse interaction mechanism in multiferroic perovskites was investigated by Sergienko and Dagotto \cite{Sergienko}. They showed that it is energetically advantageous for the negatively charged non-magnetic ion-ligand to shift from the position of the mass center to a distance perpendicular to the bond of the positively charged magnetic ions. 

The investigation of multiferroics is not restricted to obtaining static polarization. The dynamic magnetoelectric effect is an important field of research. The applied alternating electric field of an electromagnetic wave can produce a time-variable polarization $\frac{d{\bf P}}{dt}\neq 0$. Such an electric field is primarily the electric component of the electromagnetic wave in the terahertz or far-infrared range \cite{Pimenov}. Magnons caused by photo-induced oscillation of the electric polarization can be detected in multiferroics \cite{Talbayevet,Tokura2} and can lead to a dynamic magnetoelectric effect \cite{Omid}. An ultrafast way to generate and control the Dzyaloshinskii-Moriya interaction and spin currents in a class of multiferroic magnets using a terahertz circularly polarized laser and the Floquet formalism for periodically driven systems was proposed in Ref. \cite{Sato}.

There are generally accepted approaches to the magnetoelectric effect description, as noted above. But there is no complete macroscopic theory of the ferroelectricity of spin origin. For example, there is no clear understanding of the relation between the spin current and the spin-orbit interaction in the spin current model \cite{Katsura,Mostovoy,Sergienko}. Some theoretical models for describing polarization in spiral magnets are either phenomenological in nature or are also related to the study of single-particle dynamics \cite{Mostovoy,Mostovoy2}. On the other hand, control of polarization and its dynamic properties in spiral magnetic structures requires the search for new theoretical models and approaches because it's very important to understand the microscopic mechanisms affecting polarization dynamics. For these purposes, we proceed with two novel approaches for this field of research. \textit{First, we aim to propose a new theoretical model for describing the polarization in systems with dipole moments proportional to the vector product of the spins of positively charged ions}. This model must be non-phenomenological, and the main results of this model must follow from the basic principles of the method. For these purposes, we apply the method of many-particle quantum hydrodynamics for the description of ferroelectricity of spin origin, which is an effective method for investigating processes in non-equilibrium systems of many-interacting particles \cite{MQH0,MQH1,MQH2,MQH3,MQH4,MQH5,MQH6,MQH7,MQH8}. The quantum hydrodynamics method has already been applied to study nonequilibrium processes in plasmas and plasma-like media \cite{MQH0,MQH1,MQH2}, scalar and spinor Bose-condensates \cite{MQH3,MQH4,MQH5}, Fermi liquids \cite{MQH7}, superconductivity, and other fields of research \cite{MQH8}. 

The idea to use the many-particle quantum hydrodynamics method to describe the magnetoelectric effect of spin origin is new for this field of research. Therefore, it is important for us that the results of the obtained theoretical model are consistent with those derived from other methods and approaches, for example, such as the Ginzburg-Landau approach described above. \textit{Second, an important task is to investigate exactly the non-equilibrium properties of electric polarization}. At the moment, there is no model that could represent the polarization evolution equation for strongly correlated magnetic dielectrics. For example, the basic dynamical equation of magnetization evolution was introduced by Landau and Lifshitz \cite{Landau2}, which can be supplemented by a damping term \cite{Gilbert2}. In this article, we aim to obtain such an equation for the  polarization evolution from which the polarization in static and dynamic regimes can be derived.

In this article, we propose a new nontrivial theoretical description to explain the static and dynamic magnetoelectric coupling in noncollinear multiferroics with a strong spin-orbit coupling in a magnetic field. We introduce the quantum mechanical average of the spin density operator and the operator of the electric dipole moment, which is proportional to the vector product of spins. We represent macroscopic quantities such as electric polarization and spin density (magnetization) based on the main principles of the method of many-particle quantum hydrodynamics. The derivation of the polarization evolution equation is based on the introduction of the many-particle Schr\"odinger-Pauli equation with the Hamiltonian of interactions for particles with spins, which contains the effect of Zeeman energy on the magnetic moments of the medium particles. In the first step, we seek to show the influence of the magnetic field on the polarization, so we do not intentionally consider the symmetric and antisymmetric exchange interactions to highlight the result that we are interested in. From the microscopic representation of the polarization, we find the macroscopic dynamical equation of polarization over time. The proposed model can provide predictions for polarization and enables us to consider a unified picture of magnetoelectric coupling in multiferroics in the static and dynamical regimes. We also make a direct connection between the microscopic Katsura-Nagaosa-Balatsky theory and Mostovoy's phenomenological model for magnetically induced polarization.

\section{Theoretical Model}

In some kinds of type-II multiferroics with a spiral spin structure, an electric dipole moment arises in a cell or cluster of the three-site system $M_1-O-M_2$. This system consists of two neighboring positively charged magnetic ions of transition metals $M_1, M_2$ with non-collinear spins, and the non-magnetic negatively charged ion-ligand $O$, such as the oxygen ion O$^{2-}$. These magnetic ions are localized in the crystal lattice nodes, where $\bm{S}_i$ and $\bm{S}_j$ are the spins of 3$d$ electrons. The exchange interaction between the magnetic ions is driven by superexchange through the oxygen ion located between them, which can shift from the midpoint perpendicular to the bond. Such displacement of negatively charged oxygen ions from the center of mass of positive charges on transition metal ions produces an electric dipole moment $\bm{d}_{ij}=\alpha_{ij}\cdot\bm{r}_{ij}\times(\bm{S}_i\times\bm{S}_j)$, represented in Refs. \cite{Katsura, Mostovoy, Sergienko}, where $\alpha_{ij}=\alpha(|\bm{r}_{ij}|)$ is a scalar coefficient, $\bm{r}_{ij} = \bm{r}_i - \bm{r}_j$ connects two neighboring positively charged ions directed along the propagation vector of a helical structure. 

\begin{figure}[h]
\begin{minipage}[h]{0.95\linewidth}
\center{\includegraphics[width=1.0\linewidth]{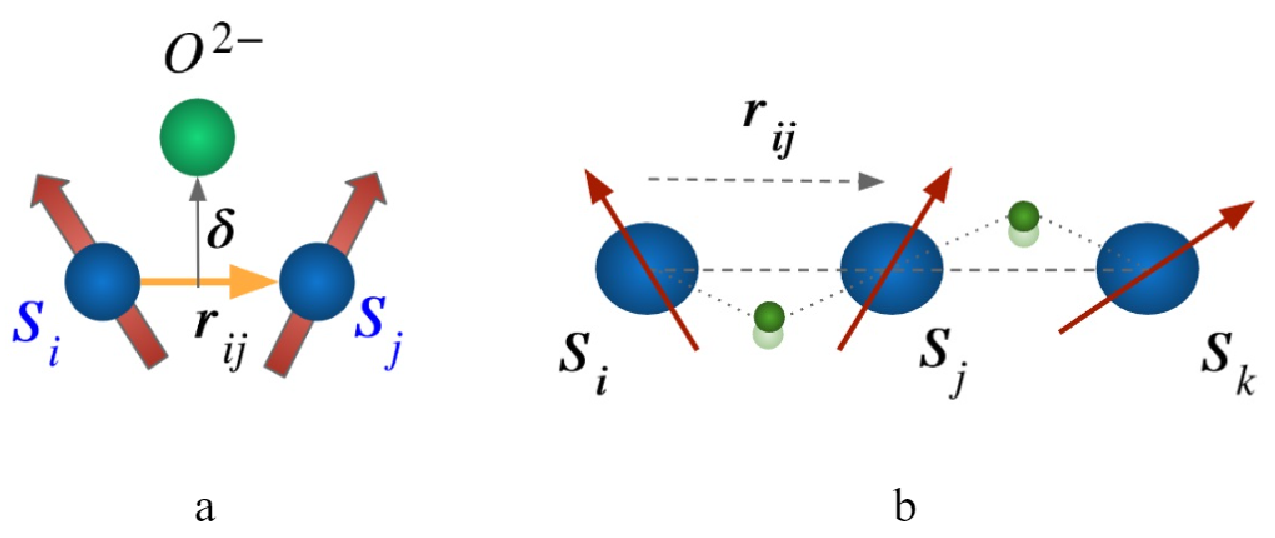}}
\end{minipage}
\caption{(Colour online) (a) - A noncollinear spin pattern (cell/cluster), which consists of positively charged metal ions (blue spheres) and a negatively charged ion-ligand (green spheres). (b) - Clockwise cycloidal spiral spin structure with oxygen displacement.}
\label{fig1}
\end{figure}

\subsection{Derivation of the polarization evolution equation}

The first step of the method is the microscopic representation of the macroscopic functions of physical fields. In a quantum mechanical description, it is necessary to find the average value of the corresponding operator of a physical quantity on the wave function of a quantum state. When the latter is an eigenstate of the physical quantity operator, the average value of the operator coincides with operator's eigenvalue. An important physical variable is the local spin density of the ions $\bm{S}$ (magnetization $\bm{M}$) and the electric dipole density $\bm{P}$ (polarization) of the medium. 

In order to introduce a macroscopic physical quantity such as the electric polarization of the medium, it is necessary to define the dipole moment operator in the vicinity of the $i$-th ion. Let us determine the \textit{operator of the electric dipole moment}, which is induced in the cell with the $i$-th magnetic ion
\begin{equation}\label{Elect}
\hat{\bm{d}}_i = \sum_{j,j\neq i}\alpha(|\bm{r}_{ij}|)\hat{\bm{r}}_{ij}
\times(\hat{\bm{S}}_i\times\hat{\bm{S}}_j).
\end{equation}
The coefficient $\alpha_{ij}$ decreases rapidly with the distance between ions $r_{ij}$. To introduce the electric dipole moment operator, we have used the definition for the electric dipole of the cell, which was derived in the spin-current mechanism \cite{Katsura}. The macroscopic polarization in the $\bm{r}$ point neighborhood in the physical space can be derived from the definition of the quantum observable as a quantum average of the operator (\ref{Elect}) of the electric dipole density 
\begin{equation} \label{Dipole}
\bm{P}(\bm{r},t)=\sum_{s}\int dR\sum_{i=1}^N\delta(\bm{r}-\bm{r}_i)\psi^{\dagger}_{s}(R,t)\hat{\bm{d}}_i\psi_{s}(R,t).
\end{equation}
In a similar manner, the magnetization of the medium is also determined as the quantum mechanical average of the operator $\hat{\bm{\mu}} = \sum_i\delta(\bm{r} - \bm{r}_i)\gamma_i\hat{\bm{S}}_i$ of the magnetic dipole moment density 
\begin{equation} \label{Magnet}
\bm{M}(\bm{r},t)=\sum_{s}\int dR\sum_{i=1}^N\delta(\bm{r}-\bm{r}_i)
\psi^{\dagger}_{s}(R,t)\gamma_i\hat{\bm{S}}_i\psi_{s}(R,t).
\end{equation}
Here $dR = \Pi_{i=1}^Nd\bm{r}_i$ is the volume element in $3N$ dimensional configuration space $R=\{\bm{r}_1,\dots,\bm{r}_N\}$ of $N$ magnetic ions, $\hat{\bm{S}}_i$ is a spin operator of $i$-th ion, $\gamma_i$ is the gyromagnetic ratio, and $s=\{s_1,\dots,s_N\}$. The microscopic definitions of the electric polarization \eqref{Dipole} and magnetization \eqref{Magnet} are the starting point for further construction of the theoretical description.

In the method of many-particle quantum hydrodynamics, we introduce a many-particle wave function of the ions' state $\psi_s(R,t)$ that depends on $3N$ spatial coordinates $R=\{\bm{r}_1,\dots,\bm{r}_N\}$ and time $t$. The many-particle wave function is a $N$-rank spinor
\begin{equation}
  \psi_s(R,t)=\psi_{s_1,...,s_N}(\bm{r}_1,...,\bm{r}_N,t),
\end{equation}
and it contains information about the quantum system of ions. Its evolution satisfies the \textit{Pauli-Schr\"odinger} time-dependent equation $i\hbar\partial_t\psi_s(R,t) = (\hat{H}\psi)_s(R,t)$, where $(\hat{H}\psi)_s(R,t) = \sum_{s'}\hat{H}_{ss'}\psi_{s'}(R,t)$. 

Next, we introduce the many-particle Hamiltonian of interactions $\hat{H}$. When studying magnetically ordered structures, it is necessary to include magnetocrystalline anisotropy, the applied magnetic field, the demagnetization field, and the exchange interactions. Polarization switching by an external magnetic field is actively investigated in experiments. Therefore, at the initial stage of theory construction, we include in the Hamiltonian only the Zeeman energy. Accordingly, we consider a system of magnetic ions in an external magnetic field that affects the spins through the Zeeman energy. In this case, the many-particle Hamiltonian of interactions has the form
\begin{equation}\label{H0}
  \hat{H}=-\sum^N_{k=1}\gamma_k\hat{S}^{\beta}_kB^{\beta}_{k},
\end{equation}
where $B^{\alpha}_{k}$ is the external magnetic field acting on the magnetic moment of the $k$-th ion. The account of exchange interactions and other types of interactions is considered separately, and a more general case with an account of the Coulomb exchange interaction, which is significant for the generation of electromagnons, as well as the Dzyaloshinskii-Moriya interaction are considered in recent Refs. \cite{https://doi.org/10.48550/arxiv.2403.01211,https://doi.org/10.48550/arxiv.2312.16321}. 

At this stage, we focus on the derivation of the dynamical equation for the evolution of the electric dipole density (electric polarization). For this purpose, we have to take the derivative of the function (\ref{Dipole}) with respect to time $\partial_t\bm{P}(\bm{r},t)$. This derivative acts on the wave function $\psi_s(R,t)$ under the integral $\partial_tP^{\alpha}=\sum_{s}\int dR\sum_{i=1}^N\delta(\bm{r}-\bm{r}_i)(\partial_t\psi^{\dagger}_{s}\hat{d}^{\alpha}_i\psi_{s}+\psi^{\dagger}_{s}\hat{d}^{\alpha}_i\partial_t\psi_{s})$. The time derivatives of the many-particle wave function $\partial_t\psi_{s}$ can be expressed through the Hamiltonian (\ref{H0}), which yields
\begin{equation}\label{EQ1}
\partial_tP^{\alpha}(\bm{r},t)=\frac{i}{\hbar}\sum_{s}\int dR\sum^N_{i}
\delta(\bm{r}-\bm{r}_i)\psi^{\dagger}_{s}[\hat{H},\hat{d}_i^{\alpha}]\psi_{s}.
\end{equation}
In order to obtain the macroscopic polarization equation from the microscopic definition of the electric dipole moment density (\ref{Dipole}), we substitute the microscopic Hamiltonian of the interactions (\ref{H0}) into the equation (\ref{EQ1}) and extract the sum
\begin{equation}\label{sumdel}
\delta^+_{ij} := \delta(\bm{r}-\bm{r}_i) + \delta(\bm{r}-\bm{r}_j)
\end{equation}
of the delta-functions of the $i$-th and $j$-th ions:
\begin{widetext}
\begin{eqnarray}
\partial_tP^{\alpha}(\bm{r},t) &=& \frac{1}{2}\sum_{s}\int dR\sum^N_{i,j\neq i}\psi^{\dagger}_{s}\delta^+_{ij}
\alpha_{ij}\epsilon^{\alpha\beta\gamma}\epsilon^{\gamma\mu\nu}\epsilon^{\delta\nu\sigma}r^{\beta}_{ij}\left(\gamma_j
B_j^{\delta}S^{\mu}_iS^{\sigma}_j-\gamma_iB_i^{\delta}S^{\mu}_jS^{\sigma}_i\right)\psi_s\nonumber\\
&=& \sum_{s}\int dR\sum^N_{i,j\neq i}\psi^{\dagger}_{s}(R,t)\delta^+_{ij}\alpha_{ij}\gamma_j
\left(\hat{S}^{\alpha}_i\cdot(\epsilon^{\beta\gamma\delta}B^{\beta}_jr^{\gamma}_{ij}\hat{S}^{\delta}_j)+(\bm{r}_{ij}
\cdot\hat{\bm{S}}_i)\epsilon^{\alpha\beta\gamma}B^{\beta}_j\hat{S}^{\gamma}_j\right)\psi_s(R,t).\label{EQ2}
\end{eqnarray}
\end{widetext}
The commutation relations for the spin  operators $[\hat{S}_{i}^{\alpha},\hat{S}_{j}^{\beta}] = i\hbar\delta_{ij}\varepsilon^{\alpha\beta\gamma}\hat{S}_{i}^{\gamma}$ and $\hat{S}^{\alpha}_i\hat{S}^{\beta}_j=\hat{S}^{\beta}_j\hat{S}^{\alpha}_i,  i \neq j$ have been used in the derivation of Eq. (\ref{EQ2}).

\subsection{Small-parameter expansion}

On the next step, we distinguish the contribution of the coordinates of the $i$-th and $j$-th ions in the delta-functions $\delta(\bm{r}-\bm{r}_i)$, $\delta(\bm{r}-\bm{r}_j)$ and wave function argument $\psi_s(R,t)=\psi_s(\bm{r}_1,\dots,\bm{r}_i,\dots,\bm{r}_j,\dots,\bm{r}_N,t)$. For these purposes, we introduce the coordinates of the center of mass
\begin{equation}
\bm{R}_{ij} = {\frac {\bm{r}_i + \bm{r}_j}{2}},
\end{equation}
and the relative motion $\bm{r}_{ij} = \bm{r}_i - \bm{r}_j$ for the $i$-th and $j$-th ions. The position vectors of the $i$-th and $j$-th ions are then expressed in terms of the center of mass and the relative distance
\begin{equation}
\bm{r}_i = \bm{R}_{ij} + \frac{\bm{r}_{ij}}{2}, \qquad \bm{r}_j = \bm{R}_{ij} - \frac{\bm{r}_{ij}}{2}.
\end{equation}
The coefficient $\alpha_{ij}$ characterizes the interactions in the cell, and decreases rapidly as the relative distance between ions $\bm{r}_{ij}$ increases. Because of this, the whole integrand in the equation (\ref{EQ2}) tends to zero at large values of the relative distance between ions. Therefore, we can expand the delta-functions, wave function and magnetic field that acts on the $j$-ion into a Taylor series. The magnetic field and the sum of delta-functions (\ref{sumdel}) are expanded up to the first and second orders, respectively
\begin{eqnarray}
B^{\beta}_j = B^{\beta}(\bm{r}_i-\bm{r}_{ij})\approx
B^{\beta}(\bm{r}_i)-\bm{r}_{ij}\partial_{\bm{r}_i}B^{\beta}(\bm{r}_i),\\
\delta^+_{ij} \approx 2\delta(\bm{r}-\bm{R}_{ij})+\frac{1}{4}r^{\alpha}_{ij}r^{\beta}_{ij}
\partial_\alpha\partial_\beta\delta(\bm{r}-\bm{R}_{ij}).
\end{eqnarray}   
The wave function entering the equation (\ref{EQ2}), after substitution of $\bm{r}_i$ and $\bm{r}_j$, has the following explicit structure of arguments $\psi_s(R,t) = \psi_s(\bm{r}_1,\dots,\bm{R}_{ij} + \frac{1}{2}\bm{r}_{ij},\dots,\bm{R}_{ij} - \frac{1}{2}\bm{r}_{ij},\dots,\bm{r}_N,t)$, and one can expand it in a Taylor series up to the first order
\begin{equation}
\psi_s(R,t)\approx \psi_s(R',t) + \frac{r^{\alpha}_{ij}}{2}\partial^R_\alpha\psi_s(R',t),   
\end{equation}
where $R' = \bm{r}_1,\dots,\bm{R}_{ij},\dots,\bm{R}_{ij},\dots,\bm{r}_N$, and $\partial^R_\alpha = \frac{\partial}{\partial R^{\alpha}_{ij}}$ is the partial derivative with respect to $\bm{R}_{ij}$.  We take into account that the i-th dipole moment is induced in the i-th cell, and neighboring cells do not affect it. Therefore, the coefficient $\alpha_{ij}$ tends to zero outside the cell.

\subsubsection{Macroscopic polarization}

Before we turn to the derivation of the evolution equation for the polarization, let us consider the relationship between the macroscopic polarization and the magnetization. In the lowest nonzero order of the expansion, the definition of polarization \eqref{Dipole} takes the form of
\begin{equation} \label{Polarization}
P^{\alpha}(\bm{r},t)=\frac{\alpha}{\gamma}\left(M^{\beta}\partial_{\beta}M^{\alpha}
- M^{\alpha}\partial_{\beta}M^{\beta}\right),
\end{equation}
where
\begin{equation}\label{alpha}
\alpha=\frac{1}{3\gamma}\int r^2_{ij}\alpha_{ij}d\bm{r}_{ij}.
\end{equation}
The integral (\ref{alpha}) is taken over the squared relative distance $r_{ij}$, and one can have an impression that it results from the second order of the perturbation expansion. However, it is actually the first order effect, since the expression for the microscopic definition of polarization already contains terms proportional to $r_{ij}$. The resulting expression \eqref{Polarization} coincides with the result of M. Mostovoy for the magnetoelectric effect in spiral magnets \cite{Mostovoy}, where the polarization was obtained with the help of the symmetry considerations for the polarization, magnetization and thermodynamic potentials. By comparing the coefficients in the expressions for polarization, we find from Ref. \cite{Mostovoy} for the constant $\alpha=\gamma\Gamma\xi_e$, where $\xi_e$ is the dielectric susceptibility in the absence of magnetism and $\Gamma$ is the undefined coefficient of proportionality from the Ref. \cite{Mostovoy} or the inhomogeneous magnetoelectric interaction coefficient \cite{Pyatakov}.

\begin{widetext}
  
\subsubsection{Polarization evolution}

After expanding the functions into the Taylor series, the right side of the equation (\ref{EQ2}) yields at zeroth order
\begin{equation}
2\sum_{s}\int dR\sum^N_{i,j\neq i}\delta(\bm{r} - \bm{R}_{ij})\alpha_{ij}\gamma_j\psi^{\dagger}_{s}(R',t)
\Pi^{\alpha}_{ij}\psi_s(R',t),
\end{equation}
and at the first order we find
\begin{eqnarray}
- 2\sum_{s}\int dR\sum^N_{i,j\neq i}\delta(\bm{r} - \bm{R}_{ij})\alpha_{ij}\gamma_j\psi^{\dagger}_{s}(R',t)
\left(\epsilon^{\beta\gamma\delta}r^{\gamma}_{ij}r^{\mu}_{ij}(\hat{S}^{\alpha}_i\hat{S}^{\delta}_j) +
\epsilon^{\alpha\beta\gamma}r^{\delta}_{ij}r^{\mu}_{ij}\hat{S}^{\delta}_i\hat{S}^{\gamma}_j\right)
\partial^i_{\mu}B^{\beta}_i\psi_s(R',t)\\
+ \sum_{s}\int dR\sum^N_{i,j\neq i}\delta(\bm{r} - \bm{R}_{ij})\alpha_{ij}\gamma_j
\left(\partial^R_\mu\psi^{\dagger}_{s}(R',t)r^{\mu}_{ij}\Pi^{\alpha}_{ij}\psi(R',t)
+ \psi^{\dagger}_{s}(R',t)r^{\mu}_{ij}\Pi^{\alpha}_{ij}\partial^R_\mu\psi_s(R',t)\right),\label{EQ3}
\end{eqnarray}
where $\Pi^{\alpha}_{ij}=\hat{S}^{\alpha}_i\bm{B}_i\cdot(\bm{r}_{ij}\times\hat{\bm{S}}_j)+\epsilon^{\alpha\beta\gamma}(\bm{r}_{ij}\cdot\hat{\bm{S}}_i)B^{\beta}_i\hat{S}^{\gamma}_j$. The integrals over $\bm{r}_{ij}$ and $\bm{R}_{ij}$ are independent, so we can explicitly separate them $dR=dR_{N-2}d\bm{R}_{ij}d\bm{r}_{ij}$. We also take into account that $\int d\bm{r}_{ij}r^{\alpha}_{ij}\alpha_{ij}=0$ and use expansion up to the first order. As a result, denoting the integration measure $d\bm{\rho}_{12} = d\bm{R}_{12}dR_{N-2}\delta(\bm{r}-\bm{R}_{12})$, we recast the equation (\ref{EQ2}) into 
\begin{eqnarray}
\partial_tP^{\alpha} &=& \alpha\gamma B^{\beta}\epsilon^{\beta\gamma\delta}\int d\bm{\rho}_{12}
\!\cdot\!\left\{\partial_{\gamma}^{R1}\!\left(\psi^{\dagger}(R_1)\hat{S}^{\alpha}_1\psi(R_1)\right)\!\left(
\psi^{\dagger}(R_2)\hat{S}^{\delta}_2\psi(R_2)\right) - \left(\psi^{\dagger}(R_1)\hat{S}^{\alpha}_1\psi(R_1)
\right)\!\partial_{\gamma}^{R2}\!\left(\psi^{\dagger}(R_2)\hat{S}^{\delta}_2\psi(R_2)\right)\right\}\nonumber\\ 
&& +\,\alpha\gamma B^{\beta}\epsilon^{\alpha\beta\gamma}\int d\bm{\rho}_{12}\!\cdot\!\left\{\partial_{\mu}^{R1}
\!\left(\psi^{\dagger}(R_1)\hat{S}^{\mu}_1\psi(R_1)\right)\!\left(\psi^{\dagger}(R_2)\hat{S}^{\gamma}_2\psi(R_2)
\right) - \left(\psi^{\dagger}(R_1)\hat{S}^{\mu}_1\psi(R_1)\right)\!\partial_{\mu}^{R2}
\!\left(\psi^{\dagger}(R_2)\hat{S}^{\gamma}_2\psi(R_2)\right)\right\}\nonumber\\
&&-\,2\alpha\gamma\partial^{\gamma}B^{\beta}\epsilon^{\beta\gamma\delta}\cdot\int
d\bm{\rho}_{12} \psi^{\dagger}(R')\hat{S}^{\alpha}_1\hat{S}^{\delta}_2\psi(R')
-2\alpha\gamma\partial^{\mu}B^{\beta}\epsilon^{\alpha\beta\gamma}\cdot\int
d\bm{\rho}_{12} \psi^{\dagger}(R')\hat{S}^{\mu}_1\hat{S}^{\gamma}_2\psi(R').
\end{eqnarray}
To simplify the formulas, we do not write down explicitly the spin $s$-indices of the wave function. In the representation of the multiplicative expansion of the wave function and for magnetic ions of the same kind, we can make use of the substitution $M^{\alpha}(\bm{r},t)M^{\beta}(\bm{r},t)=\gamma^2\int dR_{12}dR_{N-2}\delta(\bm{r}-\bm{R}_{12})\times \psi^{\dagger}(R',t)\hat{S}^{\alpha}_1\hat{S}^{\beta}_2\psi(R',t)$. After that, we derive the macroscopic equation of polarization evolution in the form
\begin{equation}\label{polarization equation}
\partial_t\bm{P} = \gamma\,\bm{P}\times\bm{B} - \alpha\left\{\left(\bm{B}\cdot(\bm{M}\times
\bm{\nabla})\right)\bm{M} + \bm{M}\cdot\left(\bm{B}\cdot(\bm{\nabla}\times\bm{M})\right)\right\}
+ 2\alpha\left\{\bm{M}\cdot\left(\bm{M}\cdot(\bm{\nabla}\times\bm{B})\right)
+ \bm{M}\times(\bm{M}\cdot\bm{\nabla})\bm{B}\right\}.
\end{equation}
This equation is the main result of the study. One can recall the similar evolution equation for the magnetization in external fields, or the Landau-Lifshitz equation. However, the equation for the evolution of electric polarization has not been reported in the literature so far. Let us discuss the physical effects contained in the equation (\ref{polarization equation}). The term on the left-hand side of the equation, $\partial_t\bm{P},$ is a partial derivative of the polarization of the medium in time. The first term on the right-hand side represents the torque $\bm{P}\times\bm{B}$ due to an external magnetic field, and it leads to the precession of the polarization vector around the direction of the applied field. A similar torque acts on magnetization from an external magnetic field in the Landau-Lifshitz equation. The second and the third terms characterize the effect of a uniform magnetic field on the inhomogeneous magnetization $\bm{M}$. If an external field is applied perpendicular to the chirality vector of the ionic spins $\bm{S}_i\times\bm{S}_j$, the spin helix will rotate until all sub-lattice moments are in the plane perpendicular to the field $\bm{B}$. As a consequence of the rotation of the spin helix, the electric polarization vector also undergoes rotation. The last two terms in the equation are of greatest interest. They characterize the behavior of the spin density in an inhomogeneous magnetic field, which can be chosen, in particular, as the field of an electromagnetic wave in the terahertz frequency range.
\end{widetext}

\subsection{Static Polarization in Spiral Magnets}

The spiral magnetic structure and the mechanism of antisymmetric exchange interaction underlie the occurrence of ferroelectricity in many multiferroics and, above all, in perovskite with an orthorhombic structure, TbMnO$_3$ \cite{Kimura}. 

Let us consider the antiferromagnetic ferroelectric phase below the temperature of the ferroelectric order $T < T_c$. In this regime, the system of magnetic ions undergoes a helicoidal state, and magnetic moments rotate in the easy $xy$-plane \cite{Mostovoy,Kimura3}. As a result, spontaneous polarization is induced along the $y$-axis. On the other hand, the direction of this polarization may be switched by changing the magnitude of an external static magnetic field \cite{Kimura2,Kimura3}. 

Let us apply our model to obtain the polarization in such a system in the external magnetic field $\bm{B}_0 = B_0\bm{e}_y$, which is applied along the $y$-axis. This field creates angular momentum by rotating the spins (magnetic moments) of the ions in the $xz$-plane. We note that the $x$, $y$ and $z$-axes correspond to the $b$, $c$, and $a$-axes of the $Pbnm$ crystal structure. In this case, the magnetization  wave state has the general form $\bm{M}(\bm{r}) = M_{0x} \cos(\bm{r}\cdot\bm{q})\bm{e}_x + M_{0z} \sin(\bm{r}\cdot\bm{q})\bm{e}_z + M_{0y}\bm{e}_y$, where the wave vector is $\bm{q} = q\bm{e}_x$ and $M_{0x}, M_{0y}, M_{0z}$ are the constant amplitudes. If the static polarization is switched by an external uniform magnetic field $\partial_t\bm{P} = 0$. In this case, the equation (\ref{polarization equation}) takes the form
\begin{equation}\label{polarization equation1}
\gamma P_x\bm{e}_z - \gamma P_z\bm{e}_x-\alpha (M_z\partial_x)\bm{M} + \alpha(\partial_xM_z)\bm{M} = 0,
\end{equation}
and leads to the non-zero solution for the polarization vector in the $z$-axis direction
\begin{equation} \label{solution}
P_z = {\frac{\alpha}{\gamma}}(M_x\partial_x M_z - M_z\partial_x M_x) = {\frac{\alpha}{\gamma}}qM_{0x}M_{0z}.
\end{equation}
As we can see, the spiral spin structure rotates in the $xz$-plane, and the non-zero polarization vector lies in the direction of the $z$-axis. The solution (\ref{solution}) confirms the result, which was derived from the definition \eqref{Polarization} and with the polarization obtained in Mostovoy's model \cite{Mostovoy} for a spiral spin-density-wave state. The macroscopic polarization \eqref{Polarization} has been here derived from the microscopic formula for electric dipole moment, which was proposed by Katsura, Nagaoshi and Balatsky, in the lowest nonzero order of the perturbation expansion. Thereby, both results \eqref{Polarization} and \eqref{solution} lead to Mostovoy's formula for the polarization and this clearly demonstrates the self-consistency of the theory. Even though the present model is incomplete and it should be extended to include the Dzyaloshinskii-Moriya interaction and Coulomb exchange explicitly, it confirms the well-known experimental result. Further development of the model is in progress to describe more complex dynamic cases and predict new dynamic and even static effects.

\section{Conclusions}

The theoretical description of the microscopic mechanisms of ferroelectricity in multiferroics is a highly nontrivial issue. The microscopic mechanism of the magnetoelectric effect based on the spin supercurrent model for a crystal bond with non-collinear spins was obtained by Katsura, Nagaosa and Balatsky \cite{Katsura}. As was shown, the electric dipole moment arises due to the spin-orbit coupling, which distorts the electron cloud that surrounds an ionic core. The induced macroscopic polarization in spiral magnets was derived by M. Mostovoy using the Landau-Lifshitz approach \cite{Mostovoy}. Both approaches give a description of static polarization. On the other hand, in contrast to the Landau-Lifshitz equation, which describes the dynamical magnetization evolution in a local magnetic field \cite{Mochizuki}, no similar description of the polarization dynamics was known in the theoretical models of ferroelectricity. In this article, we attempted to offer a new microscopic representation of the electric polarization, and a new dynamical equation was obtained for this quantity in the framework of the many-particle quantum hydrodynamics approach. This method allows us to study the non-equilibrium properties of systems with a large number of interacting ions. The resulting equation for the polarization dynamics (\ref{polarization equation}) contains only one free parameter $\alpha$, it can be used to estimate the value of ferroelectricity, and provides new ways through which magnetoelectric coupling can be understood. In accordance with the equation (\ref{polarization equation}), when an external magnetic field is applied along the $c$-axis of the crystal lattice and the cycloidal spin phase is in the $ab$-plane, the ferroelectric polarization is induced along the $a$-axis.

On the basis of the new model, we are able to establish a direct connection between the microscopic Katsura-Nagaosa-Balatsky theory and Mostovoy's phenomenological model for magnetically induced polarization. These results are also of interest for advancing the future research on dynamical magnetoelectric effects.

\begin{acknowledgments}
The research of Mariya Iv. Trukhanova is supported by the Russian Science Foundation under the grant No. 22-72-00036 (https://rscf.ru/en/project/22-72-00036/).
We would like to specially mention the fruitful discussions with Dr. Alexander P. Pyatakov.
\end{acknowledgments}

%

\end{document}